\documentclass[reprint,prl,aps,amsmath,amssymb,superscriptaddress,twocolumn]{revtex4-2}
\usepackage[utf8]{inputenc}
\usepackage[english]{babel}

\usepackage{color}
\usepackage{graphicx}
\usepackage{bm}
\usepackage{wasysym}
\usepackage{natbib}
\usepackage{fixmath}
\usepackage{comment}
\usepackage[version=4]{mhchem}
\usepackage[normalem]{ulem}
\usepackage{hyperref}
\usepackage[capitalize]{cleveref}

\def\d {{\rm d}}

\begin{document}
\title{Plastic ridge formation in a compressed thin amorphous film}

\author{Gianfranco Cordella}
\affiliation{Dipartimento di Fisica ``Enrico Fermi'', Universit\`a di Pisa, Largo B. Pontecorvo 3, I-56127 Pisa, Italy}

\author{Francesco Puosi${}^*$}
\affiliation{Istituto Nazionale di Fisica Nucleare (INFN), Sezione di Pisa, Largo B. Pontecorvo 3, I-56127 Pisa, Italy}
\email{francesco.puosi@pi.infn.it}

\author{Antonio Tripodo}
\affiliation{Dipartimento di Fisica ``Enrico Fermi'', Universit\`a di Pisa, Largo B. Pontecorvo 3, I-56127 Pisa, Italy}

\author{Dino Leporini}
\affiliation{Dipartimento di Fisica ``Enrico Fermi'', Universit\`a di Pisa, Largo B. Pontecorvo 3, I-56127 Pisa, Italy}
\affiliation{Istituto per i Processi Chimico-Fisici-Consiglio Nazionale delle Ricerche (IPCF-CNR), Via G. Moruzzi 1, I-56124 Pisa, Italy}

\author{Anaël Lema\^{\i}tre}
\affiliation{Navier, Ecole des Ponts, Univ Gustave Eiffel, CNRS, Marne-la-Vall{\'e}e, France}

\date{\today}

\begin{abstract}
We demonstrate that surface morphogenesis in compressed thin films may result from spatially correlated plastic activity. A soft glassy film strongly adhering to a smooth and rigid substrate and subjected to uniaxial compression, indeed, does not undergo any global elastic pattern-forming instability, but responds plastically via localized burst events that self-organize, leading to the emergence of a series of parallel ridges transverse to the compression axis. This phenomenon has been completely overlooked, but results from common features of the plastic response of glasses, hence should be highly generic for compressed glassy thin films.
\end{abstract}

\maketitle

 \begin{figure*}[t]
\includegraphics[height=0.55\textwidth,width=0.9\textwidth]{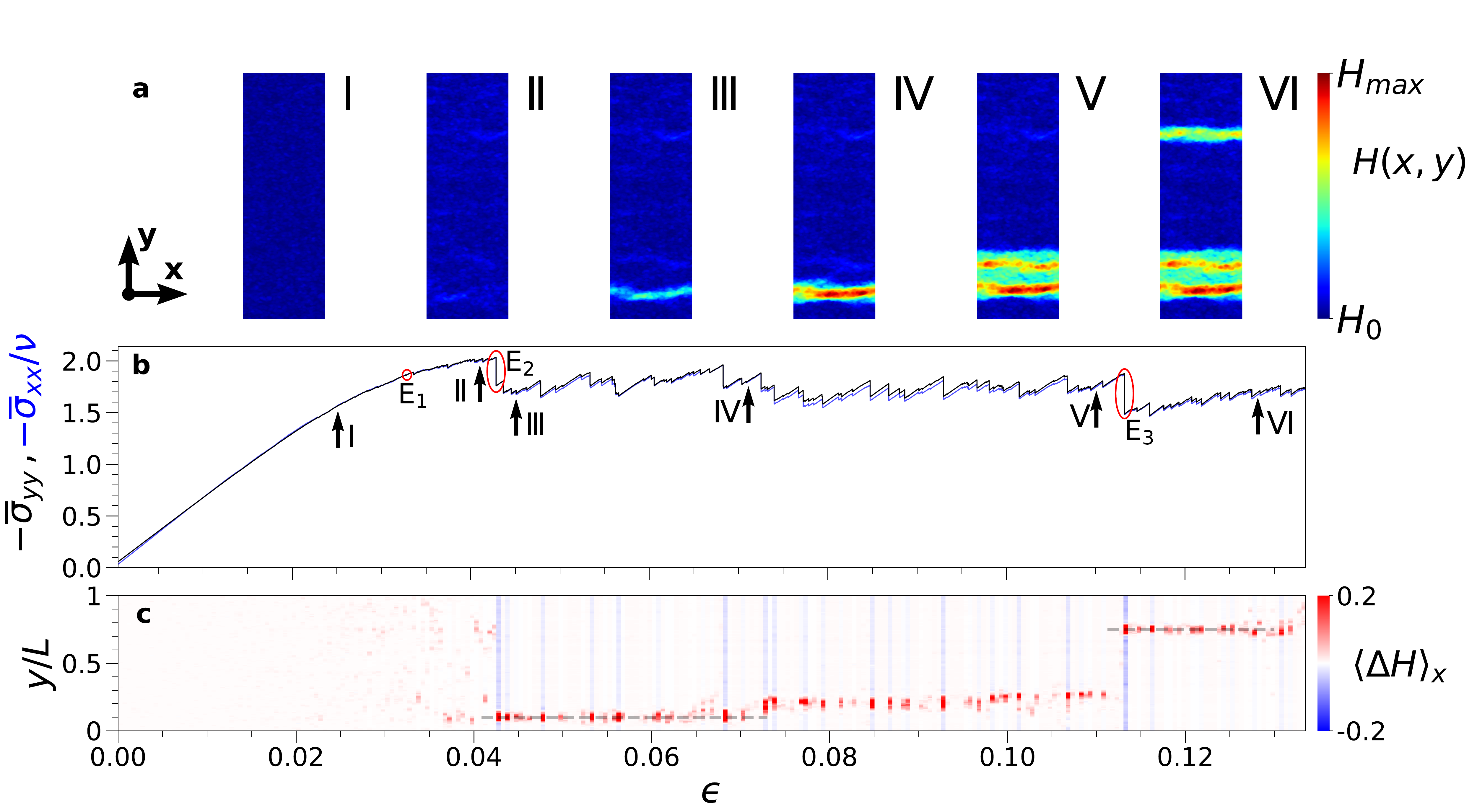}
\caption{\label{fig1} a) Height maps of our system, at different values $\epsilon$ of the uniaxial strain along $y$; each colormap is normalized between $H_0\simeq7.2$ (the initial film height) and $H_{max}\simeq13.7$  the maximum height in configuration VI . b) The mean (macroscopic) stresses $-\overline{\sigma}_{xx}/\nu$ and $-\overline{\sigma}_{yy}$ , as a function of $\epsilon$. c) The $x$-averaged height change over intervals $\Delta\epsilon = 0.0005$ as a function of $\epsilon$ and $y/L$.}
\end{figure*}

Few physical phenomena are more fascinating than morphogenesis, the spontaneous emergence of shapes in materials that are out of equilibrium as a result of, e.g., internal growth or chemical processes or in response to external driving forces. A case of primary interest, which plays a prominent role in a variety of natural and technological processes ranging from  thin film metrology~\cite{stafford_buckling-based_2004}, biological film growth~\cite{asally_localized_2012}, manufacture of smart adhesion surfaces~\cite{chan_surface_2008}, optical structures~\cite{chan_fabricating_2006,van_den_ende_voltage-controlled_2013}, hydrophobic surfaces~\cite{chung_anisotropic_2007,lin_mechanically_2009}, flexible electronics~\cite{song_mechanics_2009}, is the emergence of surface patterns in compressed thin films. In films made of soft materials or hard solids at low temperatures, this pattern formation process does not involve atomic diffusive transport: it is ``athermal'', i.e., primarily results from the mechanical response of the system composed of the film and the substrate supporting it.


Theoretical studies of pattern formation in athermally compressed films have almost exclusively focused on cases when the phenomenon results from elastic instabilities taking place well below the material yield stress. The hallmark of these studies is the wrinkling instability, which may occur at rather low strains, from a few percent to much smaller values, depending on conditions, yet is only relevant to thin films that are either stiffer than the substrate~\cite{genzer_soft_2006}, or float on a liquid~\cite{brau_wrinkle_2013}, or are freely standing~\cite{cerda_geometry_2003}. 
Homogeneous and strongly adhering films on rigid substrates do not display wrinkling, but another elastic instability called creasing, which however may only take place beyond very large strains $\epsilon \simeq 0.35$~\cite{hohlfeld_unfolding_2011} and is hence only relevant to films made of materials having a very high elastic limit, such as gels~\cite{trujillo_creasing_2008} or elastomers~\cite{hong_formation_2009}.

Meanwhile, surprisingly little attention has been dedicated to the possible role of plasticity in surface morphogenesis in compressed films. While some experiments clearly point to the occurrence of plastic strains accompanying, e.g., the growth of surface ridges or labyrinthine patterns~\cite{takei_high-aspect-ratio_2016,yang_stretching-induced_2017}, plasticity is usually viewed as a by-product of the elastic instabilities~\cite{gurmessa_onset_2013}, but not as a possible driver for the pattern formation itself.

Here, we demonstrate that surface structures may emerge solely as a result of correlated plastic activity, in the absence of any macroscopic elastic instability. This phenomenon is relevant to thin films having a low elastic limit, such as biofilms, metal or oxide layers, adhering strongly to a rigid substrate. The relevant macroscopic elastic instability, creasing, would then occur much beyond the elastic limit of the material and is thus superseded by plasticity. Using atomistic simulations we demonstrate that multiple parallel ridges transverse to the compression axis emerge as a result of the spatially correlated accumulation of burst events corresponding to a local upward motion of the material.

The emergence of multiple ridges results from a type of interrupted strain localization: (i) local bursts give rise to long-range, spatially anisotropic, Eshelby fields, which promote avalanches transverse to the compression axis; (ii) occasionally, a single massive avalanche facilitates (nucleates) strain localization, which is then sustained for a large macroscopic strain interval, thus causing the emergence of a single ridge; (iii) eventually this process stops, plastic activity abruptly delocalizes and almost immediately relocalizes elsewhere, thus leading to the emergence of another ridge. The iteration of this nonstationary process eventually gives rise to the emergence of multiple parallel ridges.\\


Molecular Dynamics simulations are performed using the atomistic model defined in~\cite{cordella_nanoscale_2021}, which consists of a set of identical, fully flexible, linear trimers, with the pair potential being harmonic for covalently bonded atoms, and Lennard-Jones (LJ) otherwise. They are implemented using the open-source software LAMMPS~\cite{LAMMPS}, and all quantities are provided in reduced LJ units. The film is rectangular and periodic in the $(x,y)$ plane, with its initial dimensions (width and length) being $(W_0, L_0)=(98,490)$ and lies in the $z>0$ half space. In order to mimic the presence of an ideally smooth and infinitely rigid substrate, a Lennard-Jones 9/3 interaction is introduced with the $z=0$ (bottom) plane. The film comprises $N= 375000$ atoms. In zero pressure conditions, it is initially contained between two walls and equilibrated at temperature $T=0.6$; after removing the upper wall, it is subjected to a cooling ramp at a rate $2\cdot 10^{-5}$ down to a very low $T$; its energy is then minimized to reach mechanical equilibrium. This produces a highly smooth (roughness $\simeq 0.06$) film of height $H_0 \simeq 7.2$. Uniaxial compression is then exerted using the athermal quasi-static (AQS) protocol~\cite{maloney_universal_2004,maloney_amorphous_2006} with adaptative strain steps~\cite{lerner_locality_2009}, i.e., by alternating homogeneous contractions along $y$, with energy minimization (using the conjugate gradient algorithm) to maintain the system at mechanical equilibrium. We denote $\epsilon=1-L/L_0$ the uniaxial strain (with $L$ the current film length), and $\sigma_{\alpha\beta}$ the $(\alpha,\beta)$ component of stress.

Due to the Poisson effect, we expect that, during the initial elastic response of the film, the mean stresses verify $\overline{\sigma}_{xx}=\nu\overline{\sigma}_{yy}$, with $\nu\simeq 0.4$ the 2D Poisson ratio. Both of these stresses (and them only), respond macroscopically to uniaxial compression, and must hence be released by the consequent plastic activity.

A typical example of film response is described in Fig.~\ref{fig1}, via height maps at different stages of deformation (panel~(a)), and plots of $-\overline{\sigma}_{xx}/\nu$ and $-\overline{\sigma}_{yy}$ vs $\epsilon$ (panel~(b)) (we checked that all other stress components fluctuate around zero, as expected). The stress plots are typical of the AQS response: they display a series of oblique branches, corresponding to the elastic response of stable atomic configurations, interrupted by stress strain-instantaneous drops, resulting from plastic events. However, the degree of coherence between the two stresses is unexpected and astonishing since $-\overline{\sigma}_{xx}/\nu$ and $-\overline{\sigma}_{yy}$ nearly perfectly collapse. We expected them to grow proportionally during elastic branches and thus to eventually relax at statistically comparable rates. But here, we clearly see that every plastic event essentially affects both of them in proportion to the Poisson ratio. It
demonstrates that a single plastic relaxation process is at work, which affects both stress components.

It must be emphasized that, although these stresses appear to plateau beyond $\epsilon\simeq0.04$, the system, by construction, cannot reach a steady state: we are examining an intrinsically transient phenomenon.

The core finding of our work is evidenced by the height maps of panel~(a): the plastic deformation of the film results in the progressive emergence of a series of ridges transverse to the compression axis. Moreover, ridges appear in sequence, one after another, which demonstrates that the plastic activity is not only spatially correlated along ridge lines, but also concentrates on the emergence of single ridges over finite macroscopic strain intervals.

\begin{figure}[!t]
\includegraphics[height=0.55\textwidth, width=0.5\textwidth]{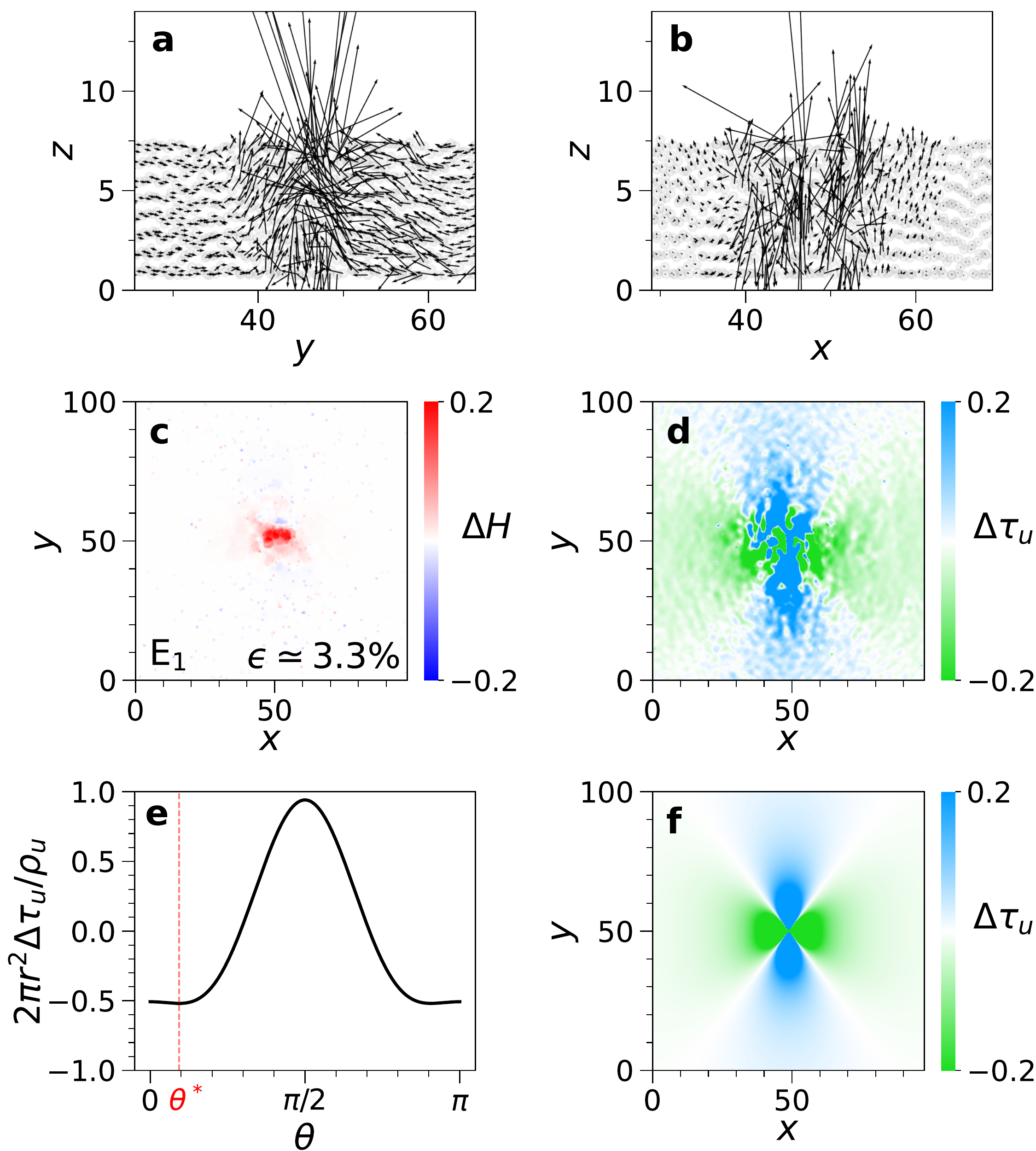}
\caption{\label{fig2} Analysis of the elementary (local) plastic event $E_1$ (see Fig.~\ref{fig1}). a,b) Cuts of the displacement field (projected in the observation plane) through the event core. c) The associated height change. d) The z-integrated stress change $\Delta\tau_{u}$ showing a stress release in blue and a strengthening in green. e) Angular dependence of the theoretical Eshelby stress change. f) Eshelby prediction for the stress change.}
\end{figure}
In order to evidence this effect, we consider the local $x$-averaged height change, $\langle\Delta H\rangle_x(y;\epsilon)$, as accumulated over slightly coarse strain steps to enhance legibility, and report it in Fig.~\ref{fig1}(c). This plot clearly demonstrates the persistence of localized plastic activity over a large macroscopic strain interval $\epsilon\in[0.043,0.071]$, which leads to the progressive emergence of a first ridge (see configurations~III and~IV in panel~(a)). 
Around $\epsilon\simeq0.072$, the first ridge stops growing; the plastic activity delocalizes over a small strain interval, before relocalizing for a large strain interval $\epsilon\in[0.073,0.111]$, thus causing the emergence of a neighboring ridge; finally, around $\epsilon\simeq 0.113$, it shifts to a completely different location, leading to the emergence of the ridge visible at high-$y$ in configuration~VI. 




The most striking and fundamental aspect of this whole process is that the plastic activity concentrates along lines transverse to the compression axis, thus enabling the formation of ridges. In order to understand why it is so, we first examine, in Fig.~\ref{fig2}, a typical highly localized event, E$_1$. From the vertical cuts of the atomic displacements (panels~(a) and~(b)) and the associated height increment map (panel~(c)), we observe that, during the event, atoms are bursting out of the compressed layer, with their displacements showing a marked $x,y$ anisotropy (compare panels~(a) and~(b)).

Why would such events self-organize into ridges? In the light of bulk plasticity studies, we expect that this might arise from the stress changes they induce in their surroundings. Taking advantage of our thin film geometry, we focus on the $\alpha,\beta=x$ or $y$ components of stress. Moreover, we note that thanks to the boundary conditions (zero stress on the upper surface and free sliding on the bottom), the integrated stress $\tau_{\alpha\beta}\equiv\int\d z\,\sigma_{\alpha\beta}$ ($\alpha,\beta=x,y$) is divergence-free, i.e., can be interpreted as a 2D stress~\cite{SI}. Additionally, to further simplify the problem, we define:
\begin{equation*}
    \begin{aligned}
    \tau_\text{u} &= \frac{\nu\tau_{xx} + \tau_{yy}}{\sqrt{2}\sqrt{1+\nu^2}}\\
    \tau_\perp &=  \frac{\tau_{xx} - \nu\tau_{yy}}{\sqrt{2}\sqrt{1+ \nu^2}}\\
    \end{aligned}
\end{equation*}
In doing so, we have diagonalized the mechanical problem, in the sense that loading now only drives $\tau_\text{u}$, but not $\tau_\perp$ which, as we observed in Fig.~\ref{fig1}(b), remains small throughout the deformation history. This enables us, from now on, to focus on $\tau_\text{u}$ only, which we call the uniaxial stress.

We display in panel~(d) the change $\Delta\tau_\text{u}$ of the uniaxial stress induced by event E$_1$ inside a square region around the event core: this field presents marked anisotropies.

As explained by Eshelby~\cite{Eshelby1957}, these stress changes are required by the elastic adaptation of the surrounding material to the local, plastic, structural reorganization taking place in the event core. Thanks to our quasi-2D geometry, we can model events using the 2D Eshelby problem in an infinite medium. In this case, a point-like plastic event occurring at the origin and locally relaxing $\tau_\text{u}$, introduces, in the surrounding medium, a change of $\tau_\text{u}$ of the form:
\begin{equation}\label{eq:eshelby}
\begin{aligned}
\Delta\tau_\text{u}  = \frac{1}{2\pi r^2} \bigg(   &-\frac{1-\nu^2}{1+\nu^2}\cos(2\theta)+\\
&\qquad+\frac{(1+\nu)(1-\nu)^2}{2(1+\nu^2)}\cos(4\theta) \bigg)  \rho_\text{u}
\end{aligned}
\end{equation}
where $\rho_\text{u}$ is the elastic dipole characterizing the event, $r = \sqrt{x^2 + y^2}$ and  $\theta = \arctan(y/x)$ .

We have numerically extracted the $\cos(2\theta)$ and $\cos(4\theta)$ components from the field $\Delta\tau_\text{u}$ induced by E$_1$ and by a few other localized events, and have thus checked that they do systematically present the expected $1/r^2$ decay. The Eshelby prediction for E$_1$ (after matching the corresponding dipole $\rho_\text{u}$), reported in Fig.~\ref{fig2}(f), demonstrates the remarkable agreement between theory and observation, modulo the expected strong near-field stress fluctuations arising from structural disorder.

The angular prefactor $2\pi r^2\Delta\tau_\text{u}/\rho_\text{u}$ vs $\theta$, plotted in Fig.~\ref{fig2}(e), is quite flat around $\theta=0$ with two minima at $\pm\theta^* \simeq 15^\circ$, thus constituting a large band of negative values around $\theta=0\,[\pi]$. This shape entails that burst events enhance (make more negative) $\tau_\text{u}$ around the transverse ($x$) axis, while attenuating it (making it more positive) along the compression direction.

From plasticity studies~\cite{maloney_universal_2004,maloney_amorphous_2006}, we know that such Eshelby stresses may induce spatial correlations of plastic activity because they bias the occurrence probabilities of future events. Here, the local stress that determines yielding probabilities is $\sigma_\text{u}$ which, thanks to our thin film geometry, we may approximate at any point $(x,y)$ by its $z$-average $\langle\sigma_\text{u}\rangle_z(x,y)=\tau_\text{u}(x,y)/H(x,y)$. Since the change in $H$ associated with a burst event is highly local, the change in $\langle\sigma_\text{u}\rangle_z$ outside the event core is $\propto\tau_\text{u}$, hence presenting the same anisotropies and long-range decay.

\begin{figure}[t]
\includegraphics[height=0.4\textwidth]{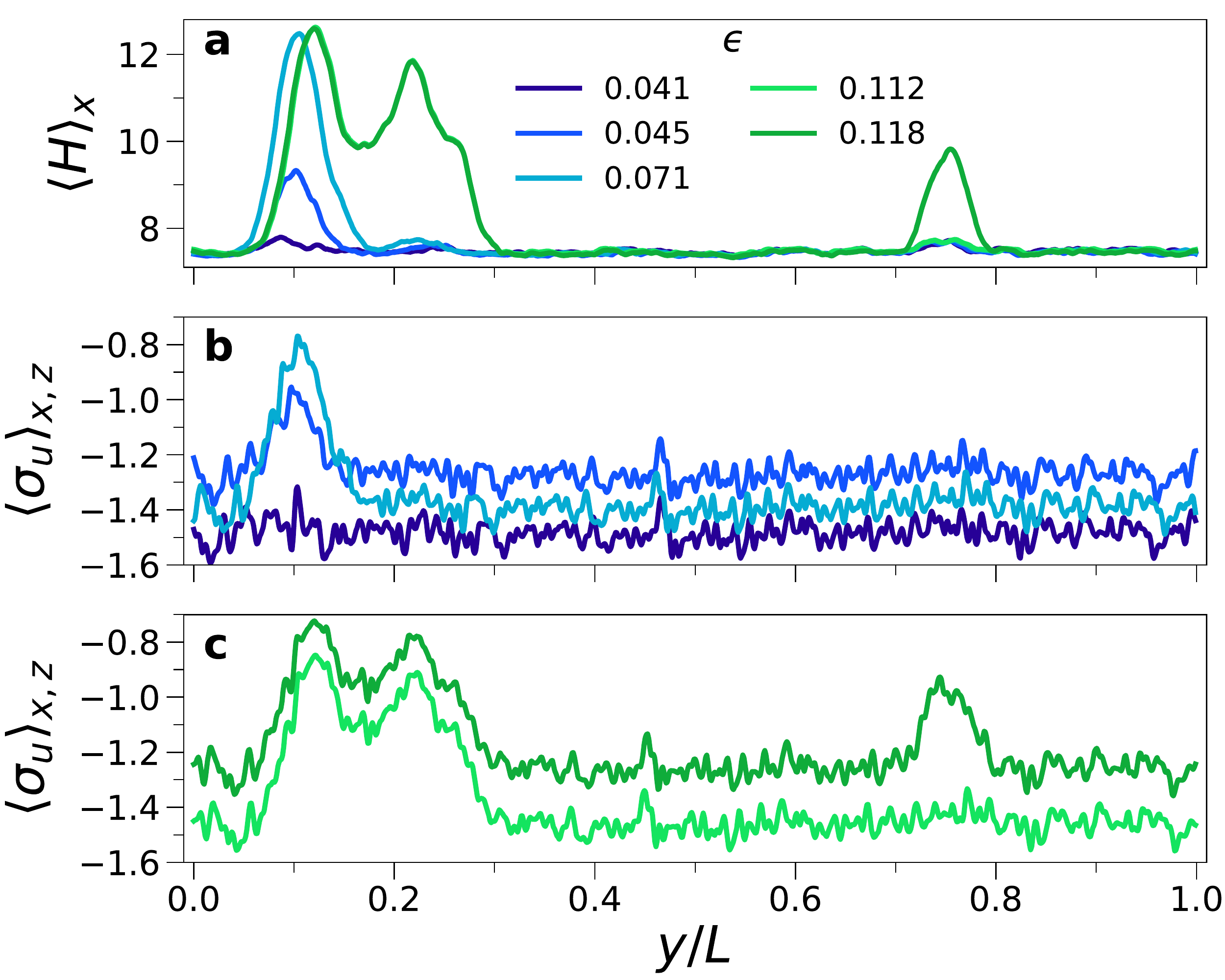}
\caption{\label{fig3} Interrupted strain localization mechanism. a) $x$-averaged height vs $y$ for several macroscopic strains [legend in panel~(a)]; b,c) $(x,z)$-averaged $\sigma_u$ vs $y/L$ for the same macroscopic strains.}
\end{figure}

In bulk plasticity, these biases are documented to induce strain-instantaneous plastic avalanches~\cite{maloney_amorphous_2006}. Here, although we cannot characterize avalanche behavior quantitatively, due to the transient nature of the ridge formation phenomenon, we found that the large stress drops display all the expected traits of avalanches: while taking place at a single macroscopic strain, they result from a series of bursts, which are strongly concentrated in $y$ and spread along the $x$ axis. Some of these events, such as E$_2$, are so massive that they span the whole cell~\cite{SI}, and lead to the partial emergence of a significant fraction of a whole ridge, as can be seen by comparing the height maps~II and~III in Fig.~\ref{fig1}. The scale of E$_2$ can be quantified from the evolution of the $x$-averaged height $\langle H\rangle_x(y)$, which is plotted in Fig.~\ref{fig3}(a): between $\epsilon\simeq 0.041$ and 0.045 (i.e., primarily during E$_2$), the first ridge grows by $\simeq 27\%$ of its final state (achieved around $\epsilon\simeq 0.071$), in terms of the volume of ejected material. The same observation can be made of event E$_3$ at strain $\approx 0.113$, which triggers the formation of the third ridge in the studied sample~\cite{SI}. We found similarly large avalanches every time activity localizes and concentrates on the emergence of a new ridge.


Two main questions remain about the phenomenology identified in Fig.~\ref{fig1}: why plastic activity concentrates around single ridges for large strain intervals, and why it abruptly delocalizes and shifts to new locations along the compression axis. For this purpose, we plot in Fig.~\ref{fig3}(b,c) the $x$- and $z$-averaged uniaxial stress field $\langle\sigma_\text{u}\rangle_{x,z}=\langle\tau_\text{u}\rangle_x/\langle H\rangle_x$ at the same strains as the height maps of panel~(a). In order to understand what determines the local value of $\langle\sigma_\text{u}\rangle_{x,z}$, let us point out that, due to mechanical balance, $\langle\tau_{yy}\rangle_x$ is strictly uniform in $y$; moreover, $\tau_\perp$ is not coupled to the uniaxial loading, hence remains small and spatially uncorrelated; as a result $\tau_\text{u}=\nu\tau_\perp+\sqrt{1+\nu^2}\tau_{yy}/\sqrt{2}$ may only present small, spatially uncorrelated fluctuations around its mean value  $\overline{\tau}_\text{u}$~\cite{SI}. Finally, $\langle\sigma_\text{u}\rangle_{x,z}\simeq\overline{\tau}_\text{u}/\langle H\rangle_x$, depends only on the total stress and on the surface profile; its $y$-dependence is hence fully determined by the height profile.

We now see that event E$_2$ takes place when $\langle\sigma_\text{u}\rangle_{x,z}$ fluctuates around -1.5 throughout the film, which provides a scale of the yield stress $\sigma_\text{u}^{y\,\text{prist}}$ in the pristine material. Right after the event, $\sigma_\text{u}$ has dropped significantly (in absolute value) inside the first ridge, compared with the rest of the system, due to the local increase in height. The observed persistence of plastic activity in this ridge implies that the material there is flowing at significantly smaller stress (in absolute value) than $\sigma_\text{u}^{y\,\text{prist}}$: this is similar to shear localization in bulk materials, and merely signals that the damage (rejuvenation) induced by the ridge inception event (and possibly its precursors~\cite{SI}) significantly reduce the local yield stress~\cite{barbot_rejuvenation_2020}.
What is highly unusual, in comparison with bulk plasticity, is that the plastic activity eventually delocalizes before rapidly relocalizing at a new location. The $\langle\sigma_\text{u}\rangle_{x,z}$ plots explain why it so happens: as height increases at an emerging ridge location, the local  $\langle\sigma_\text{u}\rangle_{x,z}\simeq\overline{\tau}_\text{u}/\langle H\rangle_x$ decreases (in absolute value).  After $\langle\sigma_\text{u}\rangle_{x,z}$ crosses under (in absolute value) the yield stress $\sigma_\text{u}^{y\,\text{dam}}\in[-1.0,-0.8]$ of the fully damaged material, the ridge growth slows down, and plastic activity eventually dies out at this location. As the film is further compressed, stress increases again in the whole system until it reaches a level $\simeq\sigma_\text{u}^{y\,\text{prist}}$ enabling strain localization in the pristine material. Here, the second ridge is triggered when $\langle\sigma_\text{u}\rangle_{x,z}$ reaches  about -1.4 (not shown); and the third (see $\epsilon=0.112$ in panel~(c)), when it reaches about -1.5.\\

In conclusion, we have identified a new mechanism of surface morphogenesis in uniaxially compressed thin films, leading to the emergence of a series of parallel ridges due to the accumulation of spatially correlated local burst events. We showed that plasticity is controlled by single stress, $\sigma_\text{u}$, which synthesizes the whole stress response to uniaxial compression and that the $\sigma_\text{u}$-releasing burst events give rise to anisotropic Eshelby fields, causing the occurrence of massive avalanches, contributing to the emergence of measurable fractions of full-blown ridges. Following a large avalanche, the plastic activity localizes in $y$, thus leading to the progressive emergence of a ridge. This process, however, eventually stops as the local increase in height causes an inversely proportional evolution of the local stress, which eventually crosses under (in absolute value) the flow stress of the damaged material. The emergence of a series of parallel ridges thus results from a form of interrupted strain localization, enabled by the contrast between the yield stresses of the pristine (undamaged) material and rejuvenated (damaged) one, and by the decay of stress at ridge locations imposed by the thin film geometry.

This mechanism of surface morphogenesis drastically contrasts with the formation of ridges, or wrinkles, arising from elastic instabilities. Yet, it appears to be rather generic, as it merely follows from commonly found features of the plastic response of glassy materials. We thus believe that it has been overlooked in the past pattern-formation literature, primarily due to the absence of a theoretical framework, and possibly due to experimental difficulties in characterizing microscale plasticity in thin films. Our findings imply that, whenever there are experimental hints of plastic activity during pattern formation, one must envision that plasticity may be a main driver for pattern formation.


\section{Acknowledgment}

FP acknowledges funding from the European Union's Horizon 2020 research and innovation programme under the Marie Sklodowska-Curie grant agreement No 754496, and from a French government grant managed by ANR within the framework of the National Program Investments for the Future, ANR-11-LABX-0022-01.

%

\end{document}


\title{Supplemental material for: ``Plastic ridge formation in a compressed thin amorphous film''}

\author{Gianfranco Cordella}
\affiliation{Dipartimento di Fisica ``Enrico Fermi'', Universit\`a di Pisa, Largo B.\@Pontecorvo 3, I-56127 Pisa, Italy}

\author{Francesco Puosi${}^*$}
\affiliation{Istituto Nazionale di Fisica Nucleare (INFN), Sezione di Pisa, Largo B.\@Pontecorvo 3, I-56127 Pisa, Italy}

\author{Antonio Tripodo}
\affiliation{Dipartimento di Fisica ``Enrico Fermi'', Universit\`a di Pisa, Largo B.\@Pontecorvo 3, I-56127 Pisa, Italy}

\author{Dino Leporini}
\affiliation{Dipartimento di Fisica ``Enrico Fermi'', Universit\`a di Pisa, Largo B.\@Pontecorvo 3, I-56127 Pisa, Italy}
\affiliation{Istituto per i Processi Chimico-Fisici-Consiglio Nazionale delle Ricerche (IPCF-CNR), Via G. Moruzzi 1, I-56124 Pisa, Italy}

\author{Anaël Lemaître}
\affiliation{Navier, Ecole des Ponts, Univ Gustave Eiffel, CNRS, Marne-la-Vall{\'e}e, France}

\date{\today}
\maketitle

\section{Problem reduction from 3D to 2D}
We want to show that our elastic problem can be reduced to a 2D problem. In doing so, it is worthwhile to do a recap of the problem, in particular its geometry and boundary conditions. Consider a box with flat faces, except the upper one in the $(x,y)$ plane with a profile given by $H= H(x,y)$. Suppose it is traction free everywhere except some normal tractions acting on the lower surface in the $(x,y)$ plane and on both surfaces in the $(x,z)$ plane. The stress inside the system $\sigma_{ij} = \sigma_{ij}(x,y,z)$, with $i$ and $j$ in $\{x,y,z\}$, must satisfy the three equilibrium equations:
\begin{equation*}
    \partial_j \sigma_{ij} =0  \;.
\end{equation*}
Now integrating each component along $z$ and writing them explicitly, one obtains :
\begin{align}\label{eq:1}
    \begin{cases}
    \intz \left( \partial_x\sigma_{xx} + \partial_y\sigma_{xy} + \partial_z \sigma_{xz}\right)  = 0  \\[10pt]
    \intz \left(\partial_x \sigma_{xy} + \partial_y\sigma_{yy} + \partial_z \sigma_{yz}\right)   = 0 \\[10pt]
    \intz \left(\partial_x \sigma_{zx} + \partial_y\sigma_{zy} + \partial_z \sigma_{zz}\right)   = 0
    \end{cases}
\end{align}
We can separate all these integrals in two classes according to whether the integrand has a partial derivative respect to $z$ or not. The first kind of integrals are of the form:
\begin{equation}\label{eq:2}
    \int_0^H dz \; \partial_z \sigma_{ij}  = \sigma_{ij}(x,y,H) - \sigma_{ij}(x,y,0) = - \sigma_{zz}(x,y,0)\; \delta_{iz} \delta_{jz}
\end{equation}
since the traction is zero on the top of the system. For the second class of integrals, we can bring the partial derivative out of the integral sign using the Leibniz integral rule: for $k=\{x, y\}$,
\begin{equation*}
\partial_k \left(\int_0^H dz \; \sigma_{ij} \right)  =  \left(\partial_k H\right) \; \partial_H\left(\int_0^H dz \; \sigma_{ij} \right)  + \int_0^H dz\; \partial_k \sigma_{ij} \; .
\end{equation*}
Thanks to the traction free conditions at the top, we furthermore observe that:
\begin{equation*}
    \partial_H \left(\int_0^H dz \; \sigma_{ij} \right)  = \sigma_{ij}(x,y,H) = 0 \quad\quad  \forall \; i,j =\{x,y,z\} \; .
\end{equation*}
So, we have obtained that in our case the following equality holds:
\begin{equation}\label{eq:3}
   \int_0^H dz\; \partial_k \sigma_{ij}  =   \partial_k \left(\int_0^H dz \; \sigma_{ij} \right)  \; .
\end{equation}
Now defining $\tau_{\alpha\beta} \equiv \int_0^H dz \; \sigma_{\alpha\beta}$, where $\alpha,\beta=\{x,y\}$, and putting together Eq.~\eqref{eq:1}, Eq.~\eqref{eq:2} and Eq.~\eqref{eq:3} we obtain :
\begin{align*}
    \begin{cases}
    \partial_x \tau_{xx} + \partial_y \tau_{xy} =0 \\[10pt]
    \partial_x \tau_{xy} + \partial_y \tau_{yy} =0 \\[10pt]
    \partial_x \left(  \int_0^H dz \; \sigma_{zx} \right) + \partial_y \left(  \int_0^H dz \; \sigma_{zy} \right) = \sigma_{zz}(x,y,0) 
    \end{cases}
\end{align*}
The first two equations can be interpreted as the mechanical balance (divergence-free condition) of a 2D problem with stress $\tau_{\alpha\beta}$.

Let us emphasize that our rigorous proof did not require any assumption regarding the height $H$ or the degree of deformation.

\section{Complementary results}
\subsection{Ridges and precursors}

\begin{figure}[!h]
\includegraphics[width=\textwidth]{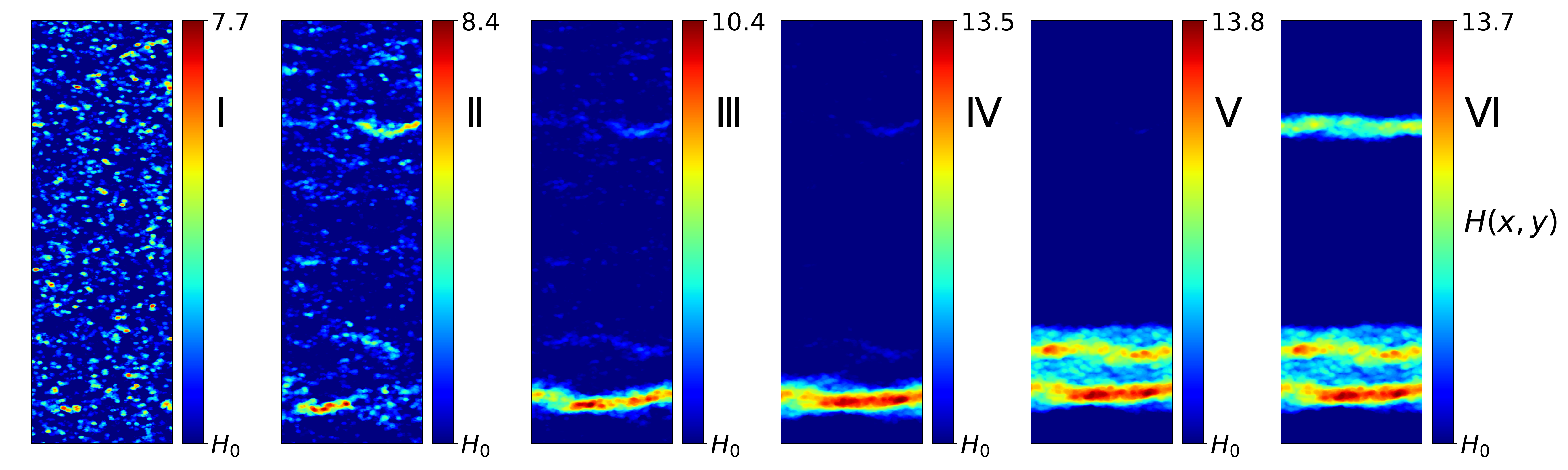}
\caption{\label{fig1_SI} Height maps of our system, at different values $\epsilon$ of the uniaxial strain along $y$; each colormap is normalized between $H_0\simeq7.2$ (the initial film height) and its own maximum.}
\end{figure}
We display in Fig.~\ref{fig1_SI} the same height maps as in the Fig.~1 in the main text, but while using a different color map for each configuration, which helps emphasizing the smaller details at the beginning of the ridge emergence process. See, in particular, a clear indication of plastic activity around the first and final ridge, already in configuration~II.

\subsection{Evidence for large avalanches}
\begin{figure}[!h]
\includegraphics[width=\textwidth]{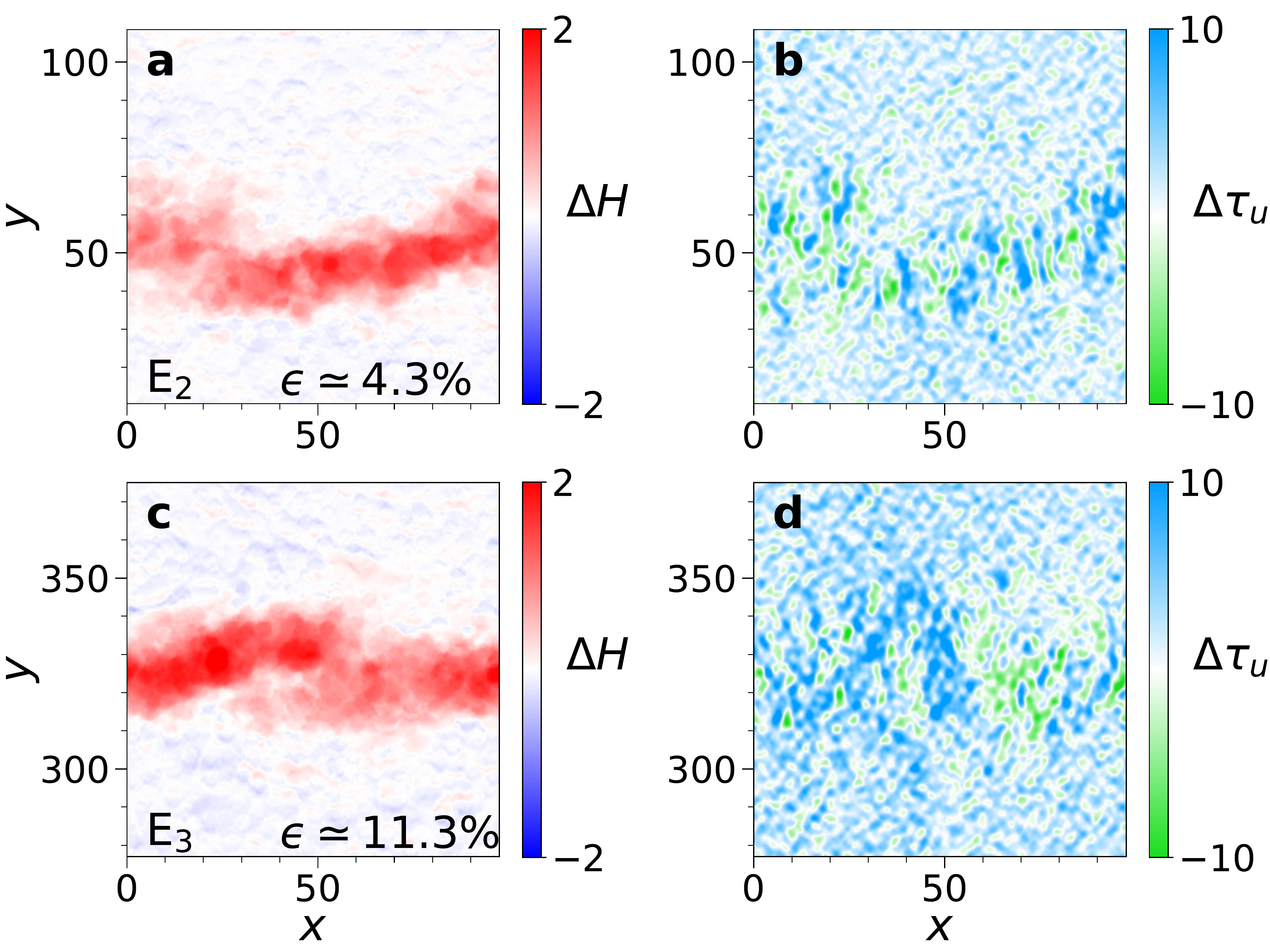}
\caption{\label{fig2_SI} Analysis of large plastic events $E_2$ (top row) and $E_3$ (bottom row) occurring at $\epsilon\simeq4.3\%$ and $\epsilon\simeq 11.3\%$ respectively. Left (panels~a,c): the height increment maps; right (panels~b,d): the z-integrated stress change $\Delta\tau_u$.} 
\end{figure}

We show in Fig.~\ref{fig2_SI} the height and stress changes associated with two large plastic events, $E_2$ and $E_3$, which occur at the onset of the first and third ridge, respectively. Both events show plastic activity spanning the whole cell transversely to the compression axis.

\begin{figure}[h!]
\includegraphics[width=\textwidth]{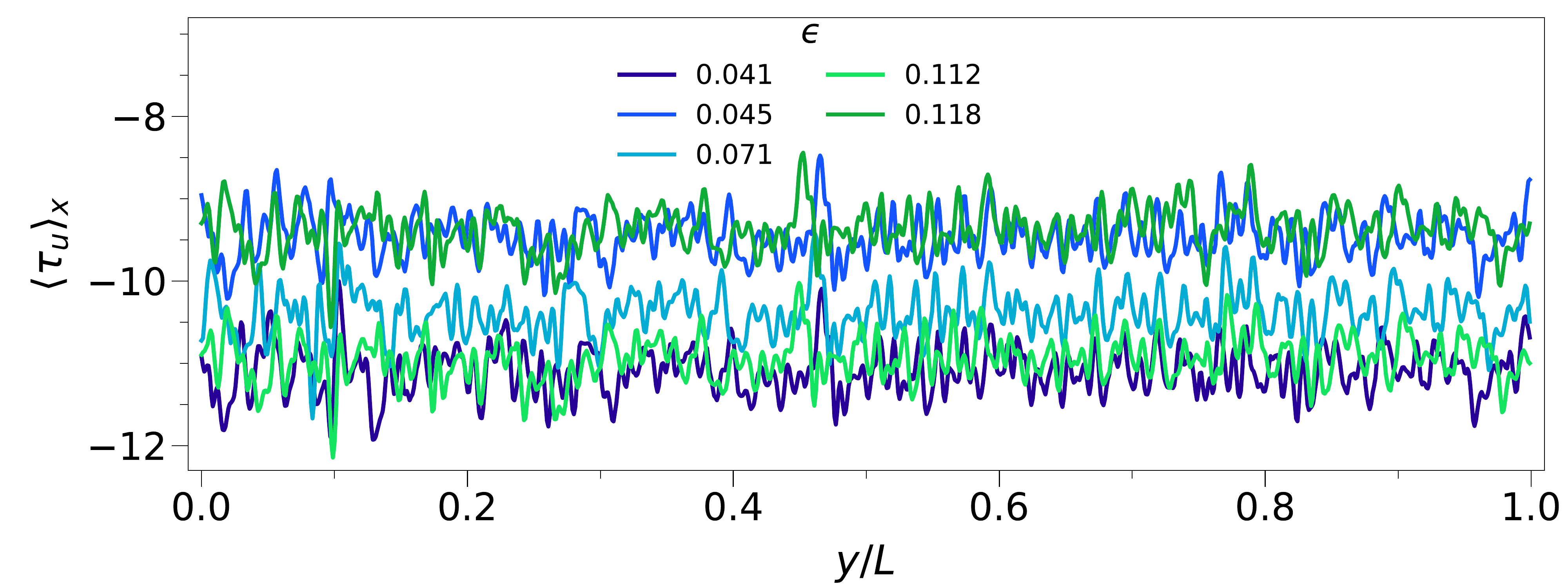}
\caption{\label{fig3_SI} Plots of $\langle\tau_\text{u}\rangle_x(y)$ vs $y/L$ for the same macroscopic strains as in Fig.~3 in the main text.}  
\end{figure}

In Fig.~\ref{fig3_SI}, we plot the evolution of $\langle\tau_\text{u}(y)\rangle_x$, i.e., the $x$-average of $\tau_\text{u}(y)$, to show that this stress field has small fluctuations, with a short spatial correlation length, and do not show any spatial correlation with the film height displayed in the main text in Fig.~3. As explained in the main text, this is a consequence of mechanical balance and of the decoupling of $\tau_{\perp}$ from the uniaxial loading.